\documentclass[aps,prd,notitlepage,amssymb,amsmath,floatfix,nofootinbib,superscriptaddress,10pt]{revtex4-1}
\usepackage{epsfig}
\usepackage{bm}
\usepackage{amssymb}
\usepackage{amsmath}
\usepackage{color}
\usepackage{slashed}
\usepackage{subfigure}
\usepackage[colorlinks,linkcolor=blue,anchorcolor=black,citecolor=blue]{hyperref}
\allowdisplaybreaks[4]

\begin{document}

\title{Quenching of polarized jets}

\author{Wen-Hao Yao}
% \email{wen-hao.yao@mail.sdu.edu.cn}
\affiliation{Institute of Frontier and Interdisciplinary Science, Key Laboratory of Particle Physics and Particle Irradiation (MOE), Shandong University, Qingdao, Shandong 266237, China}

\author{Xiaowen Li}
% \email{xiaowen.li@sdu.edu.cn}
\affiliation{Institute of Frontier and Interdisciplinary Science, Key Laboratory of Particle Physics and Particle Irradiation (MOE), Shandong University, Qingdao, Shandong 266237, China}

\author{Hui Dong}
% \thanks{Correspondence author: \href{mailto:hdong@sdu.edu.cn}{hdong@sdu.edu.cn}}
\email{hdong@sdu.edu.cn}
\affiliation{Institute of Frontier and Interdisciplinary Science, Key Laboratory of Particle Physics and Particle Irradiation (MOE), Shandong University, Qingdao, Shandong 266237, China}

\author{Shu-Yi Wei}   
% \thanks{Correspondence author: \href{mailto:shuyi@sdu.edu.cn}{shuyi@sdu.edu.cn}}
\email{shuyi@sdu.edu.cn}
\affiliation{Institute of Frontier and Interdisciplinary Science, Key Laboratory of Particle Physics and Particle Irradiation (MOE), Shandong University, Qingdao, Shandong 266237, China}

\begin{abstract}
Jets produced in association with a $Z^{0}$ or $W^{\pm}$ boson in hadronic collisions are automatically polarized due to the parity violation of weak interaction, making these processes ideal for understanding the spin transfer from polarized partons to polarized hadrons. Furthermore, leveraging this feature, we can also employ the weak-boson-tagged process to study the quenching phenomenon of polarized jets, an aspect that has rarely been discussed in the literature. In this work, we compute the polarization of $\Lambda$ hyperons in $pp$ collisions and investigate the nuclear modification due to the jet-medium interaction in $AA$ collisions. Our results highlight this process as a valuable probe of polarized parton hadronization and of spin-dependent medium effects in the QGP.
\end{abstract}

\maketitle

\section{Introduction}

The factorization theorem \cite{Collins:1981uk,Collins:1983ju,Collins:1989gx,Collins:2011zzd} allows one to separate perturbatively calculable partonic hard scatterings from the non-perturbative components of hadron production. Among these, fragmentation functions (FFs) play a central role: they describe how energetic partons hadronize into observable final-state hadrons \cite{Berman:1971xz,Metz:2016swz,Chen:2023kqw,Boussarie:2023izj}. Since a first-principles computation of FFs remains out of reach, their quantitative determination relies primarily on global analyses \cite{deFlorian:2007aj,Albino:2008fy,Aidala:2010bn,deFlorian:2017lwf,Bertone:2017tyb,Gao:2024dbv,Gao:2025bko,Gao:2025hlm} of available experimental data. Such analyses constrain the parametrization of FFs and enable predictive power in phenomenological and theoretical studies.

To incorporate the spin degree of freedom into hadron production, polarized fragmentation functions have been introduced \cite{Augustin:1978wf,Collins:1981uw,Collins:1992kk,Gustafson:1992iq,Levelt:1994np,Mulders:1995dh,Boer:1997mf,Liang:1997rt,Kotzinian:1997vd,Boros:1998kc,Ma:1998pd,Ma:1999gj,Ma:2000cg,Bacchetta:2000jk,Liu:2000fi,Bacchetta:2006tn,Collins:2007nk,Wei:2013csa,Wei:2014pma}, providing access to the spin dynamics in the hadronization process \cite{Metz:2016swz, Chen:2023kqw, Boussarie:2023izj}. In this work we focus on the longitudinal spin–transfer $G_{1L}$, which characterizes the number density of longitudinally polarized hadrons produced from longitudinally polarized partons. Although several measurements have probed $G_{1L}$ in $e^+e^-$ annihilation and polarized SIDIS \cite{ALEPH:1996oew,OPAL:1997oem,Jaffe:1996wp,deFlorian:1997kt,deFlorian:1998ba,E665:1999fso,NOMAD:2000wdf,NOMAD:2001iup,HERMES:1999buc,HERMES:2006lro,COMPASS:2009nhs,STAR:2009hex,STAR:2018pps,STAR:2023hwu}, the quantitative knowledge of this function remains limited \cite{deFlorian:1997zj,Boros:1998kc,Ma:1998pd,Ma:1999wp,Ma:2000cg,Chen:2016iey,Pan:2023pve}. Recent proposals using dihadron helicity correlations in unpolarized reactions \cite{Chen:1994ar,Zhang:2023ugf,Chen:2024qvx,Yang:2024kjn,Huang:2024awn} open a new avenue to determine polarized fragmentation functions in unpolarized colliders. 

Furthermore, relativistic heavy-ion collisions produce the strongly coupled quark gluon plasma (QGP), characterized by extremely high temperature and density \cite{Lee:1974kn,Gyulassy:2004zy,Muller:2012zq}. The interaction between high energy jets produced from hard collisions and the QGP medium results in the energy loss and the transverse momentum broadening, known as the jet quenching phenomena \cite{Gyulassy:1990ye,Wang:1992qdg, Gyulassy:1993hr,Baier:1994bd,Baier:1996kr,Baier:1996sk,Zakharov:1996fv,Gyulassy:1999zd,Gyulassy:2000fs,Gyulassy:2000er,Wiedemann:2000tf,Guo:2000nz,Wang:2001ifa,Baier:2001yt,Arnold:2002ja,Gyulassy:2003mc,Djordjevic:2003zk,Wicks:2005gt,Djordjevic:2007at,Qin:2007rn,Schenke:2009ik,Jeon:2009yv,Deng:2009ncl,Blaizot:2014bha, Iancu:2014kga, Qin:2015srf,Wang:2016fds, Barata:2020rdn, Barata:2022krd, Ghiglieri:2022gyv, Caucal:2022mpp, Adhya:2022tcn, Li:2023jeh, Mehtar-Tani:2024mvl, Barata:2025htx}. Extensive theoretical and experimental efforts over the past decades have significantly advanced our understanding of QGP properties and the mechanisms of parton–medium interactions.

While most studies have focused on momentum-space observables, the role of the spin degree of freedom in jet quenching remains largely unexplored. Only recently have investigations in this direction started to emerge. For example, Ref.~\cite{Hauksson:2023tze} examined the polarization acquired by a jet traversing an anisotropic medium. In parallel, Ref. \cite{Li:2023qgj} investigated the quenching of polarized jets utilizing the helicity correlation of back-to-back dihadron. In parallel, in the transverse-momentum-dependent factorization framework, $\Lambda$ hyperons produced within unpolarized jets also exhibit transverse polarization \cite{Belle:2018ttu, Gao:2024dxl, Matevosyan:2018jht, Gamberg:2018fwy, Anselmino:2019cqd, Anselmino:2020vlp, DAlesio:2020wjq, Callos:2020qtu, Kang:2020xyq, Boglione:2020cwn, Li:2020oto, Chen:2021hdn, Gamberg:2021iat, Yang:2021zgy, Li:2021txj, Kang:2021ffh, DAlesio:2021pxh,Kang:2021kpt,DAlesio:2021dcx,Chen:2021zrr,Ikarashi:2022yzg,Boglione:2022nzq,DAlesio:2022brl,Zaccheddu:2022qfl,Zhang:2023ugf,DAlesio:2023ozw,DAlesio:2024ope,Gao:2024bfp,Zhao:2024usu} thanks to the $D_{1T}^\perp$ fragmentation function \cite{Mulders:1995dh, Bacchetta:2004jz}. Benefiting from these progress, Ref. \cite{Qin:2025cvp} investigated the medium-induced transverse-momentum-broadening effect on the $p_T$ differential polarization. 

An alternative and complementary method of studying the hadronization of polarized partons arises in weak-boson–associated production. In unpolarized hadronic collisions, events with a reconstructed $Z^0/W^\pm$ automatically select weak-interaction processes, where parity violation produces a polarized final-state parton. This parton polarization, calculable in perturbative QCD, can be transferred to $\Lambda$ hyperons through $G_{1L}$, providing a clean probe of polarized-parton hadronization in a fully unpolarized collider environment. 

The production of the high energy jet in association with a $W^{\pm}/Z^0$ boson has been extensively studied in both $pp$ and $PbPb$ collisions at the LHC \cite{ATLAS:2012ikx,CMS:2013lua,CMS:2016php,CMS:2017eqd,CMS:2018mdf,Pellen:2022fom,Gao:2023lhs,ATLAS:2024nrd,ATLAS:2024xxl} as well as $p\bar{p}$ collisions at Tevatron \cite{CDF:2007xfq,CDF:2014sux,CDF:2014hdx,D0:2009rnw,D0:2013gro}. In weak interactions, the couplings of massive vector bosons with left-handed and right-handed particles are different \cite{Griffiths:1987tj}. For instance, in the $W$ associated quark production, it couples exclusively with left-handed quark (with helicity $\lambda_q$ = $-1$) and right handed antiquark (with $\lambda_{\bar q} = +1$). This distinct helicity structure arises from the parity violation in the weak interaction, a characteristic feature of the Standard Model. As a result, the jet produced in such interactions naturally exhibits polarization. Moreover, recent studies suggest that measuring the angular distribution coefficients of the final-state dileptons in $Z$+jet events can help distinguish between quark and gluon jet \cite{Peng:2015spa,Peng:2019boj}, which could potentially aid in extracting the flavor dependence of fragmentation functions through this process.

In this work, we compute the polarization of $\Lambda$ hyperons produced in the weak-boson tagged process in hadronic collisions and demonstrate that this observable serves as a sensitive probe to the hadronization of polarized parton. Moreover, this process also opens a novel gateway to investigate the quenching of polarized jets in heavy-ion collisions. The rest of this paper is organized as follows. In Sec.~\ref{sec:pp}, we formulate the cross section within collinear factorization to describe $\Lambda$ hyperon polarization and analyze corresponding partonic helicity amplitudes. In Sec.~\ref{sec:AA}, we detail the incorporation of energy-loss effects into the framework for $AA$ collisions. Finally, a brief summary is given in Sec.~\ref{sec:summary}.

\section{$\Lambda$ hyperon polarization in $pp$ collisions}
\label{sec:pp}

In this paper, we consider the following process in unpolarized $pp$ collisions to obtain a polarized parton jet
\begin{align}
p+p\to Z^0/W^\pm (y_V,\vec{p}_{T,V})+ \Lambda(\lambda_h, y_h,\vec{p}_{T,h})+{X},
\end{align}
where $\lambda_h$ denotes the helicity of $\Lambda$ hyperon, $y_{V/h}$ and $ \vec{p}_{T,V/h}$ are the rapidities and transverse momenta of the vector boson and the $\Lambda$ hyperon respectively. After the hard scattering, the vector boson and the parton are produced nearly back-to-back in the transverse plane, with the $\Lambda$ hyperon being produced from the fragmentation of the parton.

\subsection{Differential cross section and $\Lambda$ polarization}

The differential cross section in the collinear factorization can be expressed as \cite{Owens:1986mp}
\begin{align}
\frac{\mathrm{d}\sigma_{\lambda_h}}{\mathrm{d}y_V \mathrm{d}y_h \mathrm{d}^2\vec{p}_{T,V}\mathrm{d}^2\vec{p}_{T,h}}
= \int \frac{\mathrm{d}z_d}{z_d^2}\sum_{ab\to Vd}
\sum_{\lambda_d} x_af_{a}(x_a)\;x_bf_{b}(x_b) 
\frac{1}{\pi} \frac{\mathrm{d}\hat{\sigma}^{ab\to Vd}_{\lambda_d}}{\mathrm{d}\hat{t}} 
\delta^2 \left( \frac{\vec{p}_{T,h}}{z_d}+\vec{p}_{T,V} \right) 
\mathcal{D}_d^{\Lambda} (z_d,\lambda_d;\lambda_h) ,
\end{align}
where $z_d$ denotes the momentum fraction of parton $d$ carried by the $\Lambda$ hyperon and $\hat{\sigma}$ is the partonic cross section, with the subscript $\lambda_d$ denoting the helicity of the final-state parton and $\hat{t}$ the Mandelstam variable. $f_{a}\;(x_{a})$ and $f_{b}\;(x_{b})$ are the collinear parton distribution functions with $x_{a,b}$ representing the momentum fractions carried by partons $a$ and $b$. $\mathcal{D}_d^{\Lambda}(z_d,\lambda_d;\lambda_h) $ represents the helicity-dependent fragmentation function, which can be decomposed as \cite{Zhang:2023ugf}
\begin{align}
\mathcal{D}_d^{\Lambda} (z_d,\lambda_d;\lambda_h)=D_{1,d}^\Lambda (z_d)+\lambda_d\lambda_h G_{1L,d}^\Lambda(z_d),
\end{align}
with $ D_{1,d}^\Lambda (z_d)$ being the unpolarized fragmentation function and $ G_{1L,d}^\Lambda(z_d) $ quantifying the longitudinal spin transfer. 

The polarization of the $\Lambda$ hyperon is defined as the ratio of the difference to the sum of the cross sections with different helicities. Integrating over the transverse momenta, we arrive at
\begin{align}
\mathcal{P} _{\Lambda}(y_V,y_h) 
= \frac{\int \frac{\mathrm{d}z_d}{z_d^2} \int \mathrm{d}^2 \vec{p}_{T,h} \sum_{ab\to Vd} x_af_{a}(x_a) x_bf_{b}(x_b) \lambda_d
\frac{1}{\pi} \frac{d\sigma_U^{ab\to V d}}{d\hat t}
G_{1L,d}^{\Lambda}(z_d)}
{\int \frac{\mathrm{d}z_d}{z_d^2}\int \mathrm{d}^2 \vec{p}_{T,h} \sum_{ab\to Vd} x_af_{a}(x_a) x_bf_{b}(x_b)  
\frac{1}{\pi} \frac{d\sigma_U^{ab\to V d}}{d\hat t} D_{1,d}^\Lambda (z_d)},
\end{align}
where the unpolarized partonic cross section is obtained by summing over the helicity of final state parton $d\sigma_U = d\sigma_+ + d\sigma_-$, and $\lambda_d$ is the helicity of final state parton which can be evaluated by $\lambda_d = (d\sigma_+ - d\sigma_-)/d\sigma_U$. 

\subsection{Partonic cross section}

Now, we need to evaluate the partonic cross section for the $a(p_a)+b(p_b) \to V(p_c)+d(p_d,\lambda_d)$ process at the leading order level with the helicity amplitudes method\cite{Gastmans:1990xh}. Since we are interested in the polarized final state hadron in unpolarized collisions, we average over the spin degree of freedom of both initial-state partons, and sum over the spin indice of the final-state massive vector boson. 

\begin{figure}[htb]
\includegraphics[width=0.5\textwidth]{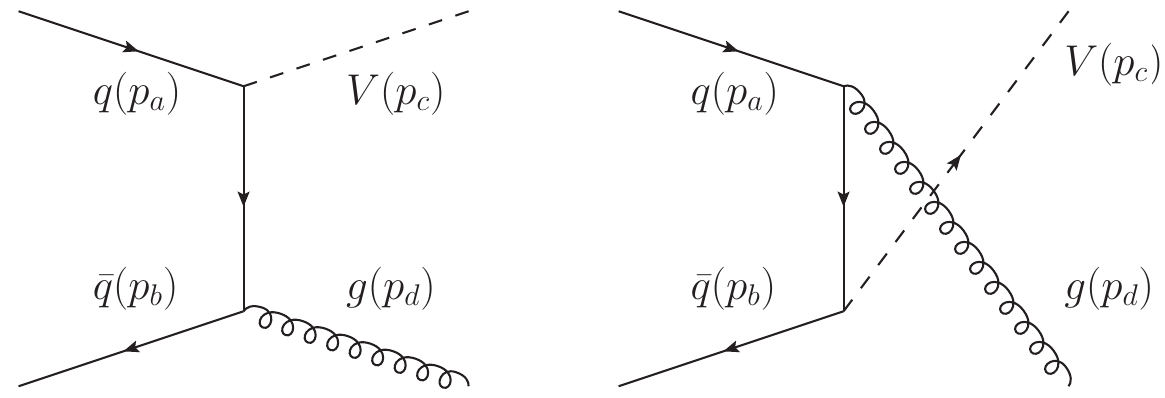}
\caption{Feynman diagrams of the $q+g \to V +q$ channel with $V$ representing a $Z^0$ or $W^\pm$ boson.}
\label{fig:fey-qqbar}
\end{figure}

The Feynman diagram of gluon production associated with a $Z^0/W^\pm$ boson is presented in Fig. \ref{fig:fey-qqbar}. For the $Z^0+g$ production, the partonic cross sections are given by
\begin{align}
&\frac{\mathrm{d}\sigma_+^{q\bar{q}\to {Z}^0 g}}{\mathrm{d}\hat{t} }
= 
\frac{4\pi \alpha_s\alpha_{\rm em}}{9 \hat{s}^2 \sin^2 2\theta_W} \left\{ c_1^q \left[ \frac{2M_Z^2 \hat s}{\hat{t} \hat{u}}+\frac{\hat{t}}{\hat{u}}+\frac{\hat{u}}{\hat{t}}\right] + c_3^q \left[\frac{2M_Z^2(\hat{t}-\hat{u})}{\hat{t}\hat{u}}-\frac{\hat{t}}{\hat{u}}+\frac{\hat{u}}{\hat{t}}\right]\right\}, \label{eq:sigma+_z}
\\
&
\frac{\mathrm{d}\sigma_-^{q\bar{q}\to {Z}^0 g}}{\mathrm{d}\hat{t} }
= \frac{4\pi \alpha_s\alpha_{\rm em}}{9 \hat{s}^2 \sin^2 2\theta_W} \left\{ c_1^q \left[ \frac{2M_Z^2 \hat s}{\hat{t} \hat{u}}+\frac{\hat{t}}{\hat{u}}+\frac{\hat{u}}{\hat{t}}\right]- c_3^q \left[\frac{2M_Z^2(\hat{t}-\hat{u})}{\hat{t}\hat{u}}-\frac{\hat{t}}{\hat{u}}+\frac{\hat{u}}{\hat{t}}\right]\right\},\label{eq:sigma-_z}
\end{align}
where $M_Z$ is the mass of $Z^0$ boson, $\theta_W$ is the Weinberg angle, and $c_1^q = (c^q_{V})^2+(c^q_{A})^2$ and $c_3^q = 2c_V^q c_A^q$ with $c_V^q$ and $c_A^q$ being quark-flavor dependent coupling constants. 

For the gluon production in association with a $W^\pm$ boson, the partonic cross sections read
\begin{align}
& \frac{\mathrm{d}\sigma_+^{q_i\bar{q}_j\to {W^{\pm}}g}}{\mathrm{d}\hat{t} }= \mathrm{V}_{ij}^2 \frac{2\pi \alpha_s \alpha_{\rm em}}{9\hat{s}^2 \sin^2\theta_W} \frac{(M_W^2-\hat{u})^2}{\hat{t} \hat{u}},
\\
& \frac{\mathrm{d}\sigma_-^{q_i\bar{q}_j\to {W^{\pm}}g}}{\mathrm{d}\hat{t} }= \mathrm{V}_{ij}^2 \frac{2\pi \alpha_s \alpha_{\rm em}}{9\hat{s}^2 \sin^2\theta_W} \frac{(M_W^2-\hat{t})^2}{\hat{t} \hat{u}},
\end{align}
where $\mathrm{V}_{ij}$ is the CKM matrix element and $M_W$ is the $W$ boson mass.

\begin{figure}[htb]
\hspace{-1.6cm} 
\includegraphics[width=0.92\textwidth]{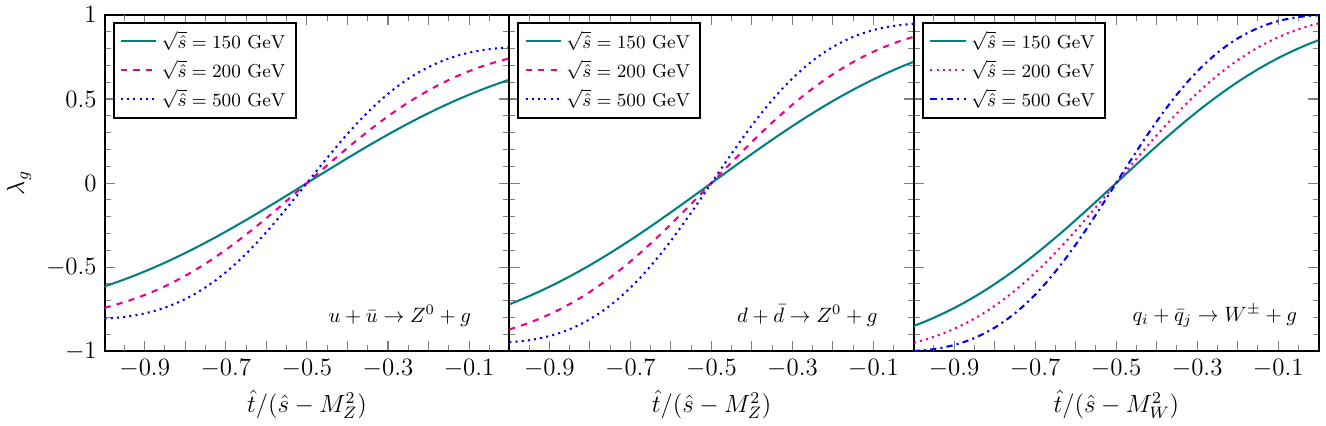}
\caption{The gluon helicity $\lambda_g$ as a function of $\hat t/(\hat s-M_{Z/W}^2)$ in $Z^0/W^\pm$ associated production processes.} 
\label{fig:gluon-helicity}
\end{figure}

It is interesting to note that the gluon helicity reverses sign under the $\hat u \leftrightarrow \hat t$ exchange. As shown in Fig.~\ref{fig:gluon-helicity}, at $\hat u = \hat t = - (\hat s - M_Z^2)/2$, the gluon polarization vanishes. The physical reason for this behavior is the following. At the $\hat t \to 0$ limit, the gluon momentum is almost parallel to that of the antiquark. Therefore, the dominant contribution for the gluon production is the collinear splitting from antiquark. In this case, the gluon helicity inherits from that of the antiquark, which is positive in weak interaction. On the other hand, if the gluon is produced in the $\hat u \to 0$ region, its helicity takes the same sign with the quark helicity which is negative. Taking into account this feature, we should avoid the mid-rapidity region with $y_V \sim y_h$ to probe the hadronization of circularly polarized gluons. Furthermore, at very small $\hat s \sim M_Z^2 \sim M_W^2$, the gluon becomes soft and carries negligible polarization. Conversely, in the high-energy limit where the $Z^0/W^\pm$ boson mass can be neglected, the gluon polarization approaches its maximum value, $\lambda_g^{\max} = \pm c_3^q/c_1^q$ for $Z^0$ boson associated process and unity for $W^\pm$, occurring at either $\hat t \sim 0$ or $\hat u \sim 0$. The gluon polarization increases monotonically with increasing $\sqrt{\hat s}$.

\begin{figure}[htb]
\includegraphics[width=0.5\textwidth]{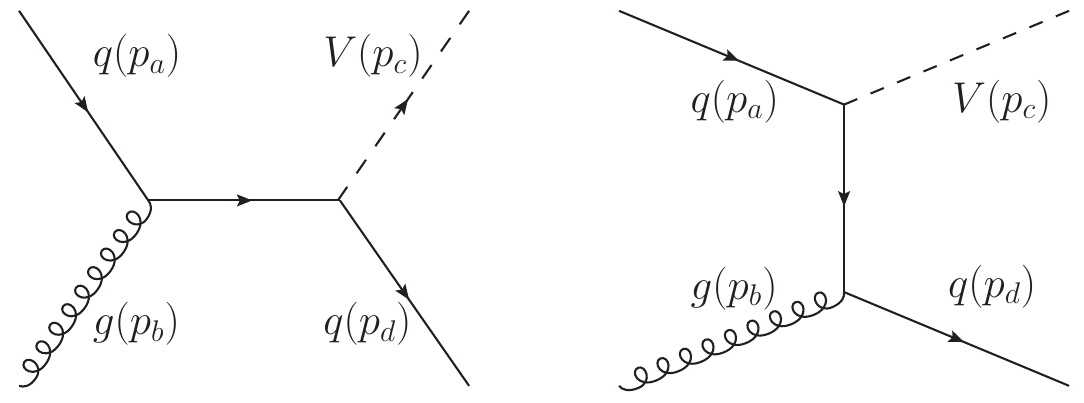}
\caption{Feynman diagrams of the $q+g \to V +q$ channel.}
\label{fig:fey-qg}
\end{figure}

The quark/antiquark production process is shown in Fig.~\ref{fig:fey-qg}, while the partonic cross sections for the $Z^0$ boson associated process are given as
\begin{align}
\frac{\mathrm{d}\sigma_+^{qg\to {Z}^0 q}}{\mathrm{d}\hat{t}} = \frac{\mathrm{d}\sigma_-^{\bar{q}g \to {Z}^0 \bar{q}}}{\mathrm{d}\hat{t} }
&
=
- \frac{\pi\alpha_s \alpha_{\rm em} }{6 \hat{s}^2 \sin^2 2\theta_W} (c_1^q - c_3^q) \left[\frac{2M_Z^2 \hat u}{\hat{t}\hat{s}}+ \frac{\hat{s}}{\hat{t}}+\frac{\hat{t}}{\hat{s}} \right],
\\
\frac{\mathrm{d}\sigma_-^{qg\to {Z}^0 q}}{\mathrm{d}\hat{t}} = \frac{\mathrm{d}\sigma_+^{\bar{q}g \to {Z}^0 \bar{q}}}{\mathrm{d}\hat{t}}
&
=
-\frac{\pi\alpha_s \alpha_{\rm em} }{6 \hat{s}^2 \sin^2 2\theta_W}  (c_1^q + c_3^q) \left[ \frac{2M_Z^2 \hat u}{\hat{t}\hat{s}}+\frac{\hat{s}}{\hat{t}}+\frac{\hat{t}}{\hat{s}} \right]. 
\end{align}

For the quark production associated with a $W^\pm$ boson, the helicity structure is unique. Only left-handed quarks and right-handed antiquarks participate in the interaction. As a result, the helicity of the final state quark/antiquark is determined, and only one helicity dependent cross section contributes which reads
\begin{align}
\frac{\mathrm{d}\sigma_-^{q_ig\to {W^{\pm}}q_j}}{\mathrm{d}\hat{t} }
=
\frac{\mathrm{d}\sigma_+^{\bar{q}_ig\to {W^{\pm}}\bar{q}_j}}{\mathrm{d}\hat{t} }
=
-\mathrm{V}_{ij}^2 \frac{\pi \alpha_s \alpha_{\rm em}}{12\hat{s}^2 \sin^2\theta_W}  \left[ \frac{2M_W^2 \hat u + \hat{s}^2+\hat{t}^2}{\hat{t}\hat{s}} \right].
\end{align}

Unlike the gluon case, the polarization of a quark or antiquark produced in association with a $Z^0$ or $W^\pm$ boson is kinematics invariant. In particular, for $W^\pm$ associated production, the polarization is always unity, with quarks carrying negative helicity and antiquarks positive. For the $Z^0$ associated production, the quark helicity is $\lambda_q = -c_3^q/c_1^q$ while the antiquark helicity reads $\lambda_{\bar q} = c_3^q/c_1^q$.

Summing over the helicities of final state particles, we arrive at the unpolarized cross sections, which coincide with those given in Refs.~\cite{Gao:2024bfp,Sun:2018icb,Field:1989uq}.

\subsection{Numerical results}

The polarized fragmentation function $G_{1L}$ is mainly extracted by analyzing the experimental data from $e^+e^-$ annihilation and polarized SIDIS process, where the gluon production is suppressed by $\alpha_s$. Therefore, our understanding on the gluon contribution remains rather limited, calling for more experimental input from hadronic colliders. 

\begin{figure}[htb]
\includegraphics[width=0.66\textwidth]{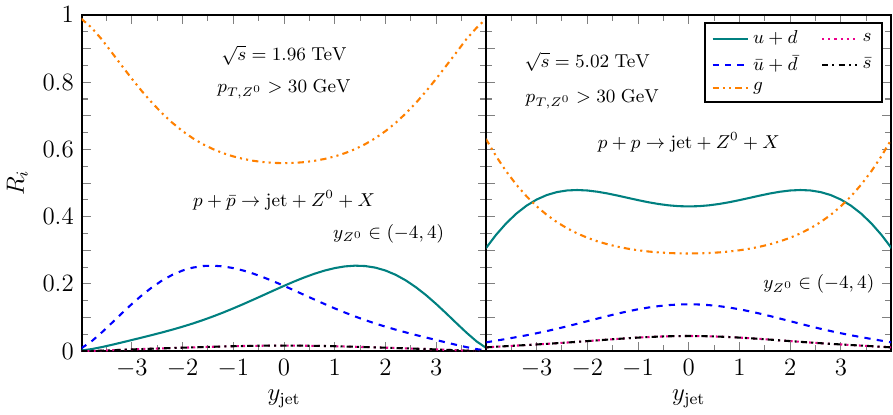}
\caption{Production ratio of different jet flavors in association with ${Z}^0$ boson in $p\bar p$ (left) and $pp$ (right) collisions.}
\label{fig:R}
\end{figure}

As shown in Fig.~\ref{fig:R}, the gluon contribution plays a significant role in the jet production associated with a $Z^0$-boson, particularly in $p\bar p$ collisions. A similar conclusion also applies to the $W^\pm$ associated production process, which is not explicitly shown. Therefore, future measurements at the Tevatron and LHC can shed more light on the hadronization of circularly polarized gluons.

\begin{figure}[htb]
\includegraphics[width=0.66\textwidth]{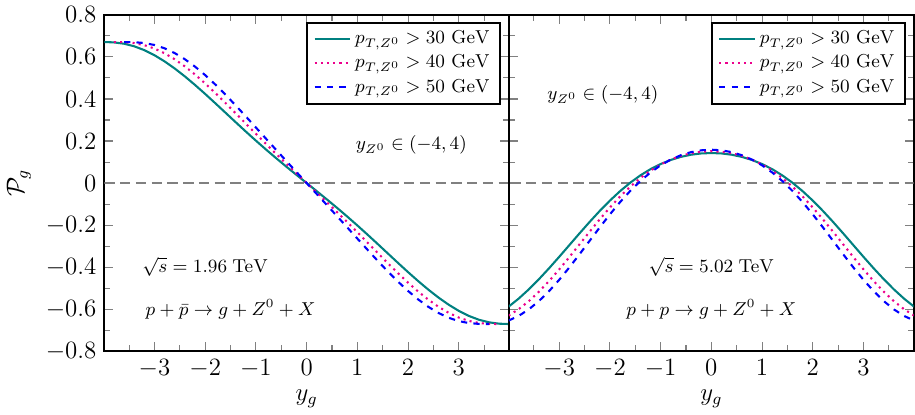}
\caption{The gluon polarization as a function of the rapidity in $p\bar p$ (left) and $pp$ (right) collisions.}
\label{fig:ppbar-g}
\end{figure}

To further quantify the gluon polarization in hadronic collisions, we compute it utilizing Eqs.~(\ref{eq:sigma+_z}-\ref{eq:sigma-_z}). The numerical results are shown in Fig.~\ref{fig:ppbar-g}, where the transverse momenta of both the $Z^0$-boson and the gluon jet are integrated over to enhance the statistical reach of potential measurements. Furthermore, we also integrate over the $Z^0$-boson rapidity and show the gluon polarization simply as a function of the gluon rapidity. 

In $p\bar p$ collisions, the gluons produced in the forward rapidity are most likely to be aligned with the quark, resulting in the negative polarization. On the other hand, the gluons in the backward rapidity are positively polarized. The magnitude of the polarization increases with the rapidity gap between the gluon and the $Z^0$-boson $|y_g-y_{Z^0}|$. In $pp$ collisions, the gluons produced in the middle rapidity also exhibit non-vanishing polarization, simply because the quark and antiquark distributions in the proton are not symmetric. In light of the sizable polarization in the forward/backward regions, hadronic colliders can indeed provide valuable insight in understanding the hadronization of circularly polarized gluons.

\begin{figure}[htb]
\includegraphics[width=0.66\textwidth]{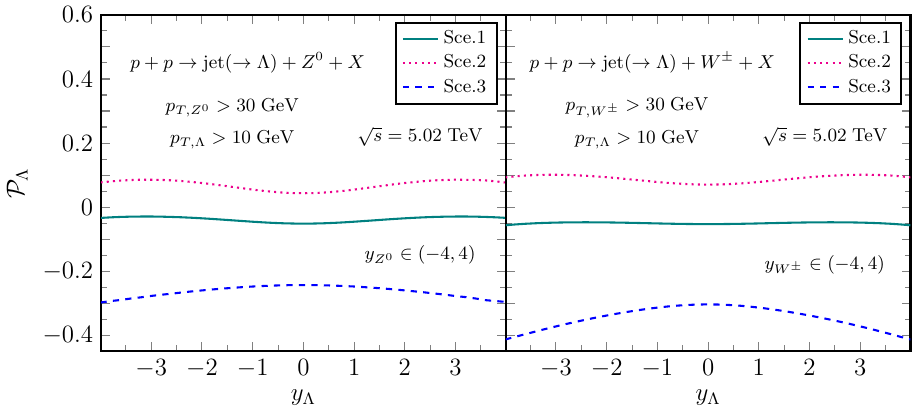}
\caption{Polarization of $\Lambda$ hyperon associated with ${Z}^0$ (left) and ${W}^{\pm}$ (right) in $pp$ collisions at the LHC energy.}
\label{fig:PL-pp}
\end{figure}

Employing the DSV parameterization of polarized and unpolarized fragmentation functions of $\Lambda$ hyperons, we make numerical predictions for the $\Lambda$ hyperon polarization in $pp$ collisions for the LHC experiment. The numerical results are shown in Fig.~\ref{fig:PL-pp}. DSV parameterization offers three flavor scenarios for the polarized fragmentation function, while all of them seem to describe the LEP data. However, in the $pp$ collisions, these three scenarios, labeled as Sce. 1-3 in Fig.~\ref{fig:PL-pp}, show distinct difference, demonstrating the potential of studying spin physics at the LHC experiment. In Sce. 1 and 2, the $s$ quark plays a dominant role in producing polarized $\Lambda$ hyperons. The contributions from the $u$, $d$ quarks and the gluon are relatively small, resulting in a very small magnitude of polarization. On the other hand, in the third scenario, $u$, $d$ and $s$ quarks are assumed to contribute equally to the $\Lambda$ polarization. Therefore, the magnitude of the $\Lambda$ hyperon polarization becomes the largest among all three scenarios. Furthermore, in Sce. 2, $u$ and $d$ quarks contribute to a small negative polarization. Therefore, The polarization in this scenario could even become positive at some kinematic region where the relative contribution from $u$ and $d$ quarks becomes more significant. Nonetheless, the significant difference among different scenarios shows that this observable is indeed sensitive to the flavor component of polarized fragmentation function. Therefore, it would be interesting to compare with the experimental data in different kinematic regions and learn more of the hadronization of polarized partons.

\section{$\Lambda$ hyperon polarization in $AA$ collisions}
\label{sec:AA}

In contrast to $pp$ collisions, heavy-ion collisions exhibit two key distinguishing features. First, the initial-state parton distributions \cite{Hou:2019efy} are modified by cold nuclear matter effects, which can be straightforwardly incorporated using available global parameterizations such as nCTEQ~\cite{Kovarik:2015cma} and EPPS~\cite{Eskola:2021nhw}. Second, the formation of the strongly coupled QGP medium induces jet quenching phenomena, characterized by parton energy loss and transverse momentum broadening. While momentum quenching has been extensively studied over the past decades, the corresponding effect on the spin degree of freedom has rarely been addressed in the literature. 

For quark jets, helicity is conserved in gluon radiation, and therefore quarks do not suffer spin quenching. The gluon case, however, is more involved. We propose two phenomenological prescriptions to account for gluon spin quenching. In the first prescription, the lost energy is carried away dominantly by a single semi-hard splitting. In this case, the helicity of the surviving hard gluon is also quenched, and the quenching magnitude can be quantified by the polarized splitting function $\Delta_L P_{gg}$, which has been computed up to two-loop accuracy \cite{Mertig:1995ny,Vogelsang:1995vh}. In the second prescription, the lost energy is carried away by multiple soft gluons, which have negligible impact on the helicity of the leading gluon. Thus, the polarization of the leading gluon remains the same while traversing the QGP medium. A detailed discussion is presented in Appendix~\ref{Spin quenching}.

\subsection{Scheme 1: Energy loss due to a single-hard-branching}

In this picture, we assume that a parton loses a small fraction of its energy through a single-hard-branching (referred to as SHB in equations and figures) and the average parton transverse energy loss $\Delta E_T$ is evaluated within the linear Boltzmann transport (LBT) model~\cite{Cao:2016gvr,Xing:2019xae,Luo:2023nsi}. Under this prescription, where heavier quarks experience less energy loss and gluons lose more energy than all quarks, we can replace the unpolarized fragmentation function in $pp$ collisions by 
\begin{align}
D_{1,d}^{\rm med} (z_d) \Bigr|_{\rm SHB} = \int \frac{\mathrm{d}\xi}{\xi^2} \xi \delta \left(\xi- \frac{k_T}{k_T + \Delta E_T} \right) D_{1,d}^{\rm vac} \left( \frac{z_d}{\xi}\right), 
\end{align}
with $k_T = p_{T,h}/z_d$ the final transverse momentum of the gluon and $\Delta E_T$ the average transverse energy loss. The normalization is determined according to the momentum conservation, which reads
\begin{align}
\sum_h \int \mathrm{d} z_d z_d D_{1,d}^{\rm med} (z_d) \approx \frac{k_T}{k_T + \Delta E_T} \sum_h \int \mathrm{d} z_d z_d D_{1,d}^{\rm vac} (z_d).
\end{align}

For quarks, helicity conservation ensures that they do not experience spin quenching while radiating gluons. For gluons, however, helicity is quenched. The suppression factor is given by the ratio of polarized to unpolarized splitting functions, $\Delta_L P_{gg}(\xi)/P_{gg}(\xi)$. At the leading order accuracy, we obtain
\begin{align}
&
G_{1L, q/\bar q}^{\rm med} (z_d) \Bigr|_{\rm SHB} = \int \frac{\mathrm{d}\xi}{\xi^2} \xi \delta \left(\xi- \frac{k_T}{k_T + \Delta E_T} \right) G_{1L,q\bar q}^{\rm vac} \left( \frac{z_d}{\xi}\right), 
\\
&
G_{1L, g}^{\rm med} (z_d) \Bigr|_{\rm SHB} = \int \frac{\mathrm{d}\xi}{\xi^2} \xi \delta \left(\xi- \frac{k_T}{k_T + \Delta E_T} \right) \frac{\xi [1-\xi(1-\xi) + (1-\xi)^2]}{[1-\xi(1-\xi)]^2} G_{1L,g}^{\rm vac} \left( \frac{z_d}{\xi}\right). 
\end{align}

\subsection{Scheme 2: Energy loss due to multiple-soft-branching}

In this scenario, the energy loss is computed within the cascade picture~\cite{Baier:1996kr, Baier:1996sk,Zakharov:1996fv}. In this model, the in-medium parton undergoes multiple-soft-branching (referred to as MSB in equations and figures) to lose a fraction of its energy. In contrast to the first scenario, since soft gluon emission does not modify the helicity of the hard gluon, it is sufficient to replace the unpolarized and polarized fragmentation functions with the nuclear modified one given by
\begin{align}
&  D_{1,d}^{\rm med} (z_d) \Bigr|_{\rm MSB}
= \int \frac{\mathrm{d}\xi}{\xi^2}  C_{dd} (\xi, \tau_{\max}) D_{1,d}^{\rm vac} (\frac{z_d}{\xi}),
\\
&
G_{1L, d}^{\rm med} (z_d) \Bigr|_{\rm MSB} = \int \frac{\mathrm{d}\xi}{\xi^2} C_{dd} (\xi, \tau_{\max}) G_{1L, d}^{\rm vac} \left(\frac{z_d}{\xi}\right),
\end{align}
where $C_{dd}$ is the momentum fraction density of parton $d$ carrying the momentum fraction $\xi$ inside the cascade initialized by parton $d$ and $\tau_{\max}$ is a dimensionless time quantifying the magnitude of jet quenching. It can be obtained by solving the in-medium parton evolution equation derived in Refs.~\cite{Blaizot:2013hx, Blaizot:2013vha, Mehtar-Tani:2018zba}.

\subsection{Numerical results} 
\label{sec:results}

\begin{figure}[htb]
\includegraphics[width=0.8\textwidth]{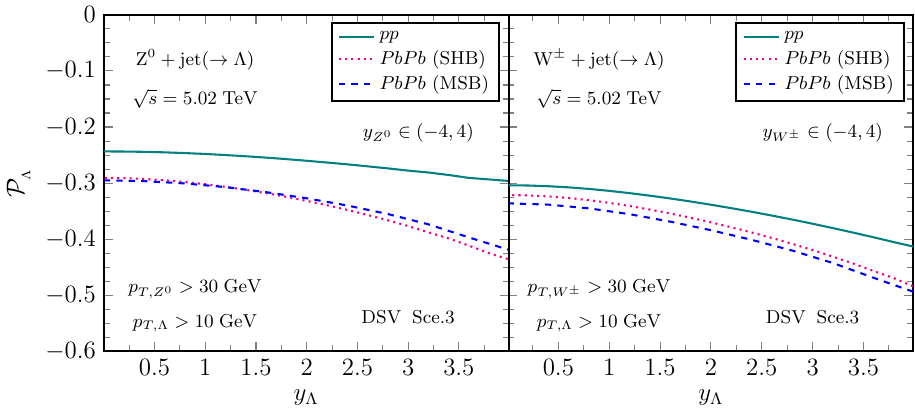}
\caption{Polarization of $\Lambda$ associated with  ${Z}^0 $ and ${W}^{\pm}$ in $pp$ and $AA$ collisions with two different schemes.}
\label{fig:ELOSS}
\end{figure}

As illustrated in Fig.~\ref{fig:ELOSS}, parton energy loss in heavy-ion collisions leads to a noticeable modification of the $\Lambda$-hyperon polarization compared with that in $pp$ collisions. This behavior can be understood as follows. Due to energy loss, the average momentum fraction $\langle z_{\Lambda} \rangle$ in the hadronization process shifts to a larger value. On top of that, the spin transfer from polarized partons to polarized hadrons is more significant at a larger momentum fraction, a feature confirmed by LEP. Therefore, the energy loss effect results in an increase of the polarization rather than a suppression. Moreover, because of the gluon spin quenching effect is very small, these two energy loss mechanisms do not result in a significant difference. Nonetheless, this observable is sensitive to the jet quenching and it would be interesting to perform a sophisticated Monte-Carlo simulation of interaction between the polarized jet and the QGP medium.

\section{Summary}
\label{sec:summary}

In this work, we focus on the weak-boson-associated jet production in hadronic collisions, and explore its potential for studying the hadronization of polarized partons. We demonstrate that the gluon polarization becomes sizable when there is a rapidity gap with the $Z^0/W^\pm$-boson, therefore paving the way for studying the {\it circularly polarized} gluons in {\it unpolarized} hadron colliders. The unpolarized colliders such as Tevatron and LHC have accumulated enormous experimental data, whose analysis could substantially reduce the ambiguities in the hadronization models and reveal vital information on the spin transfers. Moreover, this observable is also sensitive to the jet quenching in relativistic heavy-ion collisions. We investigate two representative schemes of the parton energy loss, namely the single-hard-branching scheme and multiple-soft-branching scheme. In the SHB case, the gluon loses a small portion of its polarization, while its polarization remains unquenched in the MSB. Since the total fractional energy loss is modest, the spin quenching effect in the SHB is also negligible, leading both schemes to produce similar nuclear modifications to the final-state hadron polarization.

\section*{Acknowledgments}

This work is supported by the National Science Foundations of China under Grant No. 12405156 and No. 12175118, the Shandong Province Natural Science Foundation under grant No.~2023HWYQ-011 and No.~ZFJH202303, and the Taishan fellowship of Shandong Province for junior scientists.

\appendix

% \section{Constants}
% \label{Parameters}
% The constants are given as follows
% \begin{align}
% & \alpha_{\rm em}=1/137, &&\sin^2\theta_W =0.231,
% && M_Z =91.2 \; \mathrm{GeV}, && M_W = 80.4 \; \mathrm{GeV}, \\
% & c_V^{u} = \frac{1}{2} - \frac{4}{3} \sin^2 \theta_W, && c_A^{u} = \frac{1}{2},
% && c_V^{d} = -\frac{1}{2} + \frac{2}{3} \sin^2 \theta_W, && c_A^{d} = -\frac{1}{2}.
% % & c_A^{u}=0.519, && c_V^{u}=0.266,\\
% % & c_A^{d}=-0.527, &&c_V^{d}=-0.38.
% \end{align}

\section{Spin quenching of gluon jets}
\label{Spin quenching}

While traversing the QGP medium, a circularly polarized gluon loses a fraction of its energy through medium-induced gluon radiation. At the same time, its polarization is also quenched, with the magnitude of this suppression depending on the specific parton energy-loss mechanism. To make this connection more transparent, we first recall the polarized and unpolarized gluon splitting functions which read
\begin{align}
& 
P_{gg} (\xi) = 2N_c \left( \frac{1-\xi}{\xi} + \xi(1-\xi) +  \frac{\xi}{(1-\xi)_+}\right) + \frac{11N_c - 2n_f}{6} \delta (1-\xi), \\
&
\Delta_L P_{gg} (\xi) = N_c \left( \frac{1+\xi^4 - (1-\xi)^3}{\xi} +  \frac{1+\xi^4}{(1-\xi)_+}\right) + \frac{11N_c -2n_f}{6} \delta (1-\xi).
\end{align}
The ratio $\Delta_L P_{gg} (\xi)/P_{gg} (\xi)$ characterizes the survival probability of the gluon’s polarization after one splitting that loses energy fraction $1-\xi$.  

As an illustration, consider a gluon of initial energy $E$ with helicity $\lambda_g=1$. It loses a total energy $\Delta E$ through $n$ successive splittings with each branching carrying away a small fraction of its energy. Assuming that after each branching the retained momentum fractions are approximately the same, we then obtain the per-branching retained fraction as $x = (1 - \Delta E/E)^{1/n}$. After $n$ splittings, the remaining polarization is then given by
$\lambda_g^{\rm after}/\lambda_g^{\rm before} = \left[\Delta_L P_{gg}(x)/P_{gg}(x) \right]^n$.

\begin{figure}[htb]\centering
\includegraphics[width=0.45\textwidth]{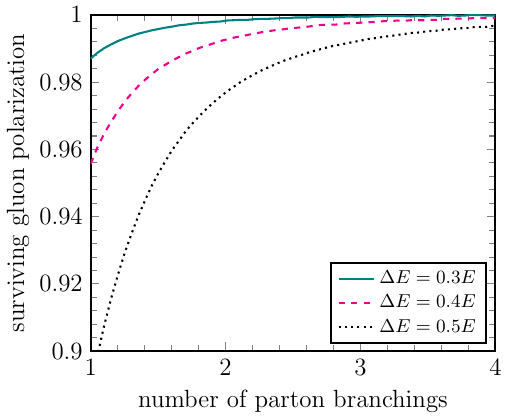}
\caption{The surviving gluon polarization after successive medium-induced branchings as a function of the number of parton branchings.}
\label{fig:rp}
\end{figure}

The numerical results are shown in Fig.~\ref{fig:rp} which exhibits two intricate features. First, the suppression is more pronounced with large total energy loss, reflecting the general property of polarized splitting function. However, the polarization loss is in general small. Even a $40\%$ energy loss only results in a $4\%$ polarization loss. Second, the surviving gluon polarization increases monotonically with the number of medium-induced branchings. This behavior arises because the energy loss per splitting becomes smaller when the total energy loss $\Delta E$ is distributed over more branchings, so each individual splitting has a weaker depolarizing effect. Consequently, the overall polarization quenching is reduced for large number of branchings $n$, and the gluon polarization is nearly preserved in the MSB limit.

To summarize, the polarization loss of the gluon while traversing the QGP medium depends on energy loss mechanisms. In the single-hard-branching energy loss picture, the polarization loss is quantified by $\Delta_L P_{gg} (1-\Delta E/E)/P_{gg} (1-\Delta E/E)$. On the other hand, in the multiple-soft-branching picture, the polarization loss becomes negligible.

\end{document}